\documentclass[
reprint,
superscriptaddress,amsmath,amssymb,aps,prd,notitlepage,longbibliography,floatfix,nofootinbib]{revtex4-1}

\usepackage{color}
\usepackage[
colorlinks=true,        
citecolor=blue,         
linkcolor=blue,         
urlcolor=blue           
]{hyperref}
\usepackage{graphicx}

\newcommand{\Eq}[1]{Eq.~\eqref{#1}} 
\newcommand{\Fig}[1]{Fig.~\ref{#1}}

\begin{document}
\title{Observational Test of the Maximum-Entropy Conjecture for Black-Hole Mergers with GW250114}

\author{Yang Jiang}
\email{jiangy@ucas.ac.cn}
\affiliation{School of Fundamental Physics and Mathematical Sciences, 
  Hangzhou Institute for Advanced Study, UCAS, Hangzhou 310024, China}
\affiliation{School of Physical Sciences, 
    University of Chinese Academy of Sciences, 
    No. 19A Yuquan Road, Beijing 100049, China}
\author{Qing-Guo Huang}
\email{corresponding author: huangqg@itp.ac.cn}
\affiliation{School of Fundamental Physics and Mathematical Sciences, Hangzhou Institute for Advanced Study, UCAS, Hangzhou 310024, China}
\affiliation{School of Physical Sciences, 
  University of Chinese Academy of Sciences, 
  No. 19A Yuquan Road, Beijing 100049, China}
\affiliation{Institute of Theoretical Physics, Chinese Academy of Sciences, Beijing 100190, China}

\date{\today}

\begin{abstract}
The remnants of binary black-hole (BBH) mergers are accurately described by numerical relativity, 
but the physical principle governing the selection of the final states remains unclear. A recently proposed maximum-entropy conjecture suggests that a BBH merger produces the Kerr BH that maximizes the Bekenstein-Hawking entropy, subject to the post-Newtonian formalism or balance laws for energy and angular momentum. In this work, we test this conjecture with the event GW250114, inferring the progenitor and remnant properties from the observed pre-merge and post-merge signals, separately. We construct the surrogate Kerr entropy associated with candidate remnant states along the inferred energy-angular-momentum trajectory. We then compare the maximum-entropy state with the independently inferred mass and dimensionless spin of the final BH. We find that the conjecture predicts a remnant mass of $M_f=70.0^{+0.35}_{-0.21}\,M_\odot$ and spin of $\chi_f=0.70^{+0.00}_{-0.01}$, compared with the ringdown estimates $M_f={72.3}^{+3.6}_{-3.7}\,M_\odot$ and $\chi_f=0.74^{+0.05}_{-0.07}$. The entropy deviation is consistent with zero within the $68\%$ credible interval for our fiducial analysis, and variations in the pre-merger truncation time and quasi-normal modes (QNM) content reveal no compelling discrepancy with the conjecture.
\end{abstract}

\maketitle

\section{Introduction}

The observation of gravitational waves (GWs) from binary black-hole (BBH) coalescence has opened a direct window onto the nonlinear and dynamical regime of gravity~\cite{LIGOScientific:2016aoc,PhysRevD.108.024043,Watarai:2025hsb}. 
Within general relativity, the inspiral stage of a BBH can be accurately modeled using post-Newtonian (PN) theory~\cite{Lorentz1937,PhysRevLett.70.2984,PhysRevD.49.2658,Blanchet:2013haa}, 
black-hole perturbation theory~\cite{Teukolsky1974,Sasaki:2003xr,Berti:2025hly}, and the effective-one-body formalism~\cite{PhysRevD.59.084006,Damour:2012mv}. 
The merger itself, however, is a genuinely strong-field process whose faithful description generally requires numerical relativity (NR)~\cite{Pretorius:2005gq,Campanelli:2005dd}. 
Large catalogs of NR simulations now allow the complete gravitational waveform, as well as the mass and angular momentum of the remnant, to be predicted across a broad region of the BBH parameter space~\cite{Scheel_2025,Ficarra:2026cej}. 
For example, an equal-mass, initially nonspinning binary on a quasicircular orbit is predicted to leave a Kerr BH with approximately $95.16\%$ 
of the initial total mass and dimensionless spin $\chi_f\simeq0.69$~\cite{Baker:2006yw,Hinder:2007qu}, 
with the remaining energy and angular momentum carried away by gravitational radiation.

Conventionally, the remnant is viewed as inheriting the energy and angular momentum remaining near plunge and subsequently losing additional amounts through merger and ringdown. 
Although NR predicts the remnant properties with remarkable precision, 
it does not by itself identify a simple physical principle that determines why a particular final state is selected. 
In contrast, BH thermodynamics may offer a natural setting in which to address this question. 
Ref.~\cite{Rincon-Ramirez:2026tbo} recently proposed a maximum-entropy conjecture for BBH mergers. 
Consider the mass $M$ of the system, including the binary binding energy, as a function of its angular momentum $J$, 
with their relation determined by the relevant energy and angular-momentum balance laws~\cite{Friedman:2001pf,LeTiec:2017ebm}. 
At each point along the trajectory of evolution, one may associate the binary with a hypothetical Kerr black hole having the same $M$ and $J$, 
and hence assign it the corresponding Kerr entropy. 
The conjecture states that the physical remnant is the Kerr configuration that maximizes this entropy along the energy-angular-momentum trajectory.
When the relation $M(J)$ is calculated using PN theory for nonspinning, quasicircular binaries, 
the mass and spin evaluated at the maximum-entropy point approach the NR predictions as higher PN orders are included.
At 4PN order, the predicted remnant properties agree with NR-based fits to within a few percent over a wide range of mass ratios.
It may therefore point to an unexpected connection between binary dynamics and BH thermodynamics.
Consequently, an observational test is  needed to determine whether this behavior is realized in nature.

The GW event GW250114~\cite{LIGOScientific:2026sit,LIGOScientific:2025wao} provides an especially favorable opportunity for such a test. 
Its exceptionally large signal-to-noise ratio (SNR) enables precise measurements of both the progenitor binary and the remnant BH~\cite{LIGOScientific:2025rid}. 
Moreover, GW250114 is consistent with a nearly equal-mass binary with small component spins and no measurable orbital eccentricity. 
These properties closely match the class of nonspinning, 
quasicircular systems for which the maximum-entropy conjecture has been most thoroughly formulated and tested with PN theory and NR. 
The strong post-merger signal also permits an independent characterization of the remnant through ringdown spectroscopy. 
Indeed, separate analyses of the pre-merger and post-merger portions of GW250114 have already provided stringent tests of Hawking's area law and of the Kerr nature of the remnant~\cite{LIGOScientific:2025rid}. 
These features make GW250114 not merely a precise test of general relativity, 
but also a unique laboratory for investigating whether a thermodynamic selection principle governs the outcome of a BBH merger.

In this work, we use GW250114 to perform an observational test of the maximum-entropy conjecture. 
We analyze the inspiral signal to infer the progenitor parameters and the binary energy-angular-momentum relation, 
from which we determine the posterior distribution of the maximum-entropy state. 
Independently, we analyze the ringdown signal to infer the mass and dimensionless spin of the remnant. 
This separation allows the maximum-entropy prediction and the observed final state to be compared without imposing the conjecture as a prior relation. 
We quantify their consistency using the posterior distribution of a deviation parameter,
establishing the first observational constraint on the formulation of the conjecture.

\section{maximum-entropy conjecture}
We begin by reviewing the maximum-entropy conjecture and formulating the quantitative prediction that will later be confronted with the pre- and post-merger analyses of GW250114.
Within the PN framework, the conservative dynamics of a compact binary are described by an expansion in a small dimensionless parameter associated with the characteristic orbital velocity~\cite{Blanchet:2002av}.
For a quasicircular binary, both the mass  $M$, including the binding energy, 
and the orbital angular momentum $J$ decrease monotonically as energy and angular momentum are carried away by gravitational radiation.
Consequently, the orbital evolution can be parametrized by $J$,
allowing the PN relation between the mass and angular momentum to be written as $M=M(J)$.
For a nonspinning binary, the gauge-invariant energy-angular-momentum relation through 4PN is~\cite{PhysRevD.89.064058}
\begin{multline}
  M = m_0 - \frac{\nu m_0}{2j^2}\Bigg\{
    1 + \left(\frac{9+\nu}{4}\right)\frac1{j^2} + \frac{81-7\nu+\nu^2}8\frac1{j^4} \\
    + \left[\frac{3861}{64} + \left(\frac{41\pi^2}{32} - \frac{8833}{192}\right)\nu - \frac{5\nu^2}{32} + \frac{5\nu^3}{64} \right]\frac1{j^6}  \\
    + \bigg[\left(\frac{6581\pi^2}{512} - \frac{989911}{1920} - \frac{128}{5}\left(\gamma_E+\ln\frac4j\right)\right)\nu  \\
    + \left(\frac{8875}{384} - \frac{41\pi^2}{64}\right)\nu^2 - \frac{3\nu^3}{64} + \frac{7\nu^4}{128} + \frac{53703}{128}\bigg]\frac1{j^8}  
  \Bigg\},
\end{multline}
where $\nu$ is the symmetric mass ratio, $m_0$ is the total initial mass and $j=J/\nu m_0^2$ is the dimensionless reduced angular momentum.

For a dynamical BBH spacetime, particularly during the highly nonlinear merger, 
there is no unique and generally accepted definition of the instantaneous entropy of the entire binary system. 
The maximum-entropy construction therefore does not identify the expression below with the actual entropy of the binary.
Instead, at every point along the PN trajectory $M(j)$, 
one considers a hypothetical Kerr black hole that would form instantaneously while inheriting the mass $M(j)$ and angular momentum $J(j)$ of the binary. 
We refer to this hypothetical object as a candidate or surrogate remnant.
The horizon area of a Kerr BH is
\begin{equation}
  A(M,J) = 8\pi M^2\left(1+\sqrt{1-\chi^2}\right),
  \label{eq:area}
\end{equation}
where $\chi=J/M^2$ is the dimensionless spin parameter.
Accordingly, the Bekenstein-Hawking entropy assigned to the candidate remnant at a given value of $j$ is~\cite{PhysRevD.7.2333}
\begin{equation}
  S(j) = \frac{A\left[M(j),\,J(j)\right]}{4\hbar}.
  \label{eq:entropy}
\end{equation}
As the binary loses angular momentum, the candidate Kerr entropy generally increases, 
reaches a maximum at a particular value $j_*$, and then decreases if the PN trajectory is formally continued beyond that point. 
The maximum-entropy conjecture states that the physically realized remnant corresponds to this maximum:
\begin{equation}
j_*\equiv\underset{j}{\operatorname{argmax}}\,S(j).
\end{equation}
Therefore, it predicts the final mass and angular momentum
\begin{equation}
M_{f}^{\rm i} = M(j_*),
\quad
J_{f}^{\rm i} = \nu j_* m_0^2,
\end{equation}

This formulation naturally separates the observational test into two parts. 
The pre-merger data determine the progenitor masses and thereby the maximum-entropy prediction $(M_f^{\rm i},\,\chi_f^{\rm i})$, 
whereas the post-merger data provide an independent measurement of the actual remnant parameters $(M_f^{\rm r},\,\chi_f^{\rm r})$ from the ringdown signal. 
The conjecture can then be tested by comparing these two posterior distributions, 
without imposing the maximum-entropy relation on the post-merger inference.

\section{Pre-merge and post-merge analyses}
GW250114 was observed by the LIGO Hanford and LIGO Livingston detectors with an exceptionally large network SNR of $\rho_{\rm net}=78.6$~\cite{LIGOScientific:2025rid}. 
The source is consistent with a nearly equal-mass BBH with small component spins and negligible orbital eccentricity. 
These properties make GW250114 particularly suitable for testing the maximum-entropy conjecture formulated previously,
since the quantitative PN relation used in our analysis is derived for nonspinning binaries on quasicircular orbits. 
The high SNR also allows the pre-merger and post-merger portions of the signal to be analyzed separately while retaining sufficient information to constrain the progenitor and remnant parameters.

We obtain the reference merger time and sky location from the inspiral-merger-ringdown (IMR) analysis of the complete signal. 
We define the geocentric merger time $t_{\rm m}$ as the time at which the GW strain, 
integrated over the celestial sphere, reaches its maximum. 
This definition is independent of the orientation of the source and coincides with the coalescence time convention adopted by the NRSur7dq4 waveform model~\cite{Varma:2019csw}. 
We fix $t_{\rm m}$ to the maximum likelihood value of the IMR posterior:
\begin{equation}
t_{\rm m}= 1420878141.235932,
\end{equation}
and fix the sky location to
\begin{equation}
({\rm RA},\,{\rm DEC})=(2.333,\,0.190).
\end{equation}
Fixing these quantities provides a common temporal and geometrical reference for the independent pre- and post-merger analyses. 
In particular, it ensures that the truncation boundaries applied to the Hanford and Livingston data correspond to the same source-frame stage of the coalescence. 
It also reduces correlations between the intrinsic parameters and weakly constrained extrinsic parameters.
Subsequently, we denote the corresponding time for detector $I$ by putting a superscript.

We analyze the portion of the signal preceding the merger using the \textsc{TDInf} package~\cite{Isi:2020tac,Miller:2025eak}.
The analyzed segment starts at $1420878140.2$ and is truncated at
\begin{equation}
t_{\rm i}^{I} = t_{\rm m}^{I} + \delta t_{\rm i}, \qquad \delta t_{\rm i}\leq0,
\end{equation}
so that only data satisfying $t<t_{\rm i}^{I}$ contribute to the pre-merger likelihood. 
Here, $\delta t_{\rm i}$ denotes the truncation time relative to the reference merger time. 
In~\cite{LIGOScientific:2025rid}, the authors choose a series of truncation times to test the robustness of the analysis.
We inherit the operation and also set the fiducial truncation time to 
\begin{equation}
\delta t_{\rm i}=-40t_M,\quad t_M=0.354\;{\rm ms}.
\end{equation}
This value excludes the two loudest signal cycles near the merger.
The strain is analyzed over the frequency interval $20\sim2038$ Hz. 
The \textsc{TDInf} code assigns  uniform priors to the total mass, mass ratio and luminosity distance.
The pre-merger waveform is modeled using NRSur7dq4; however, we do not consider the spins and fix them to 
zero in the prior. This choice corresponds to the maximum-entropy hypothesis without the effect of spin. It may introduce systematic errors in the estimation of other parameters. However, we have checked that including the estimation of spin parameters does not affect the main conclusions of this paper.

From each posterior sample of the detector-frame component masses,
we construct the 4PN relation $M(j)$ and numerically locate the entropy maximum.
This transformation maps the pre-merger posterior into a posterior distribution for the remnant predicted by the maximum-entropy conjecture
\begin{equation}
p
\left(
M_f^{\rm i},
\chi_f^{\rm i}
\mid d_{\rm i}
\right),
\end{equation}
where $d_{\rm i}$ denotes the truncated pre-merger data. The corresponding predicted entropy is calculated with \Eq{eq:area} and \Eq{eq:entropy}.

In general relativity, the ringdown of a perturbed Kerr BH can be represented as a superposition of quasinormal modes (QNMs).
The complex mode frequency is
\begin{equation}
  \tilde{\omega}_{\ell mn}=\omega_{\ell mn}(M_f, \chi_f) + i/\tau_{\ell mn}(M_f, \chi_f),
\end{equation}
where the oscillation frequencies $\omega_{\ell mn}$ and damping times $\tau_{\ell mn}$ are determined by 
the remnant mass and dimensionless spin, 
assuming that the remnant is described by the Kerr metric.
We independently infer the remnant properties from the post-merger signal using the \textsc{pyRing} package~\cite{Carullo:2019flw}. 
It assigns uniform priors to the final mass, spin and amplitude of the right- and left-handed polarized contributions to the mode.
The ringdown analysis begins at
\begin{equation}
t_{\rm r}^{I} = t_{\rm m}^{I} + \delta t_{\rm r},
\qquad
\delta t_{\rm r}\geq0,
\end{equation}
where the fiducial start time is
\begin{equation}
\delta t_{\rm r}=10.5t_{M_f},\quad t_{M_f}=0.337\; {\rm ms}.
\label{eq:Delta_tr}
\end{equation}
An earlier start time contains more signal power but may not be adequately described by a linear superposition of QNMs. A later start time is more securely within the linear ringdown regime but yields broader posterior distributions 
because of the rapidly decaying signal. 
To investigate waveform systematic effects and possible contamination from nonlinear merger dynamics, 
we repeat the analysis using different $\delta t_r$.
For a truncated start time not earlier than \Eq{eq:Delta_tr}, the significance of the $221$ mode drops below $1\sigma$~\cite{LIGOScientific:2025rid}.
Therefore, we consider the $220$ mode only in the fiducial analysis.
For every analysis configuration, \textsc{pyRing} yields the posterior
\begin{equation}
p_{\rm r}
\left(
M_f^{\rm r},
\chi_f^{\rm r}
\mid d_{\rm r}
\right),
\end{equation}
where $d_{\rm r}$ denotes the post-merger data. We use these samples to calculate the remnant Kerr entropy.

To quantify the agreement between the maximum-entropy prediction from the pre-merger data and the independently measured remnant from the post-merger data, we introduce the fractional deviation parameters
\begin{align}
\delta M_f = \frac{M_f^{\rm i}-M_f^{\rm r}}{M_f^{\rm r}},
\
\delta \chi_f = \frac{\chi_f^{\rm i}-\chi_f^{\rm r}}{\chi_f^{\rm r}},
\
\delta S = \frac{S_f^{\rm i}-S_f^{\rm r}}{S_f^{\rm r}}.
\label{eq:fracdiff}
\end{align}
In principle, the maximum-entropy conjecture predicts
\begin{equation}
\delta M_f = \delta \chi_f = \delta S = 0.
\label{eq:delta0}
\end{equation}
But the surrogate-entropy construction is approximate and need not reproduce the NR remnant exactly.
Ref.~\cite{Rincon-Ramirez:2026tbo} reports that the higher-order PN maximum-entropy predictions agree with NR fits for the remnant mass and dimensionless spin to within $1.3\%$ and $5.6\%$, respectively.
The NR-based maximum Kerr surrogate entropy differs from the final remnant entropy by at most $0.61\%$, while the 4PN-based surrogate entropy differs from its NR-based counterpart by up to $2\%$ for an equal-mass, nonspinning binary.
These comparisons characterize distinct modeling discrepancies and do not provide an event-specific error bound for our test. In particular, the quoted spin discrepancy is comparable to the statistical uncertainty reported below. We therefore adopt \Eq{eq:delta0} as a reference of exact agreement, rather than an exact null prediction free of theoretical uncertainty.

We draw samples from the pre-merger and post-merger posterior distributions to evaluate the deviation posterior. Because the Kerr entropy is completely determined by $M_f$ and $\chi_f$, 
the three deviation parameters are correlated and should not be interpreted as three independent tests. 
We use $(\delta {M_f},\delta {\chi_f})$ as the primary two-dimensional consistency diagnostic and 
$\delta S$ as a physically motivated summary of the disagreement in the thermodynamic quantity central to the conjecture. 
Finally, we repeat the consistency test for every combination of the pre-merger truncation time, post-merger start time, and QNM model. 
A genuine inconsistency with the maximum-entropy conjecture should persist across the analysis choices for which both the inspiral and ringdown models provide adequate descriptions of the data. 
In contrast, a deviation that appears only for very late pre-merger cutoffs, very early ringdown start times, 
or insufficient QNM content would be more naturally attributed to waveform-model systematics or contamination from the excluded merger regime.

\section{Results}
\begin{figure}[ht]
  \centering
  \includegraphics[width=\columnwidth]{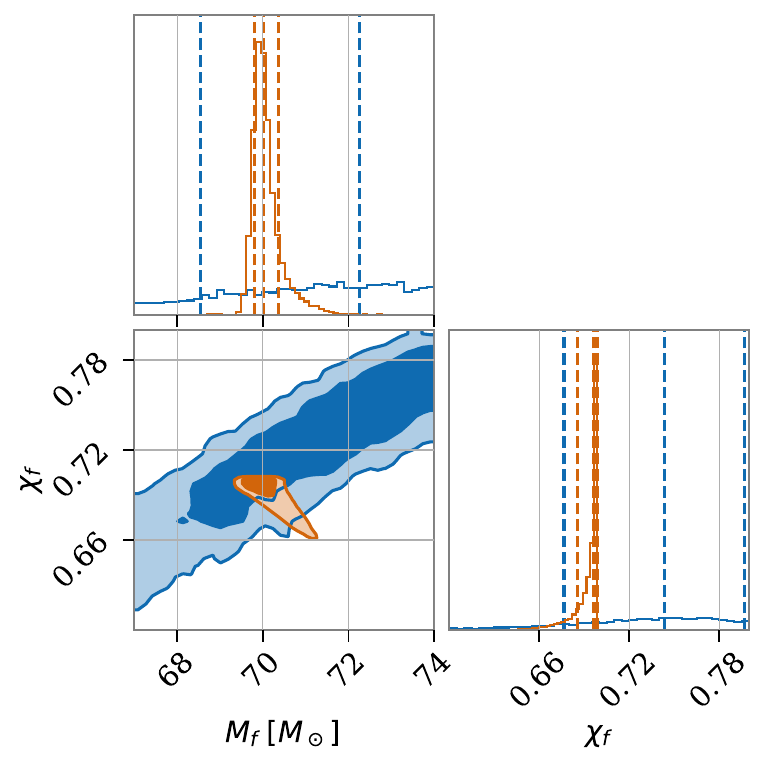}
  \caption{Posterior distributions of the mass $M_f$ and spin $\chi_f$ of the remnant black hole. 
  The orange distribution is the maximum-entropy prediction obtained by propagating the pre-merger posterior, 
  while the blue distribution is inferred from the ringdown signal. 
  The dark and light shaded contours enclose the $68\%$ and $95\%$ credible regions, respectively.
  The dashed lines denote medians and $68\%$ credible intervals.}
  \label{fig:mf-chif}
\end{figure}

We report all of the masses in the detector frame. \Fig{fig:mf-chif} compares the remnant parameters predicted from the pre-merger signal under the maximum-entropy conjecture with those inferred independently from the post-merger ringdown. 
The orange contours show the posterior distribution of $(M_f^{\rm i},\chi_f^{\rm i})$, 
obtained by propagating the inspiral posterior samples through the 4PN maximum-entropy construction. 
The blue contours show the ringdown posterior for $(M_f^{\rm r},\chi_f^{\rm r})$. 
The dark and light shaded regions enclose the $68\%$ and $95\%$ credible regions, respectively.
Note that the distribution of $\chi_f$ exhibits a sharp upper cutoff near $0.70$, with posterior probability accumulating close to this boundary. This behavior follows from the adopted nonspinning 4PN construction: the predicted dimensionless spin depends only on the symmetric mass ratio and increases monotonically with $\nu$ over the parameter domain considered. Since $\nu\leq 1/4$, the equal-mass limit sets the upper bound. The preference of the pre-merger posterior for nearly equal masses therefore produces the accumulation near this limit.
In the analysis of GW250114, the maximum-entropy analysis produces a comparatively narrow posterior centered at
\begin{equation}
M_f^{\rm i}= 70.0\,M_\odot,
\qquad
\chi_f^{\rm i} = 0.70,
\end{equation}
with substantially smaller uncertainties than those obtained from the ringdown-only analysis.
This difference in precision is expected: the pre-merger signal contains a large number of inspiral cycles and tightly constrains the binary masses, 
whereas the post-merger signal is short-lived and has a considerably lower SNR.
The ringdown posterior displays a pronounced positive correlation between $M_f^{\rm r}$ and $\chi_f^{\rm r}$.
Most importantly, 
the maximum-entropy prediction almost lies within the broad $95\%$ credible region obtained from the post-merger signal, and the two $68\%$ regions also overlap. 

We quantify the agreement using the fractional deviations $\delta {M_f}$ and $\delta {\chi_f}$ defined in \Eq{eq:fracdiff}.
\begin{figure}[ht]
  \centering
  \includegraphics[width=\columnwidth]{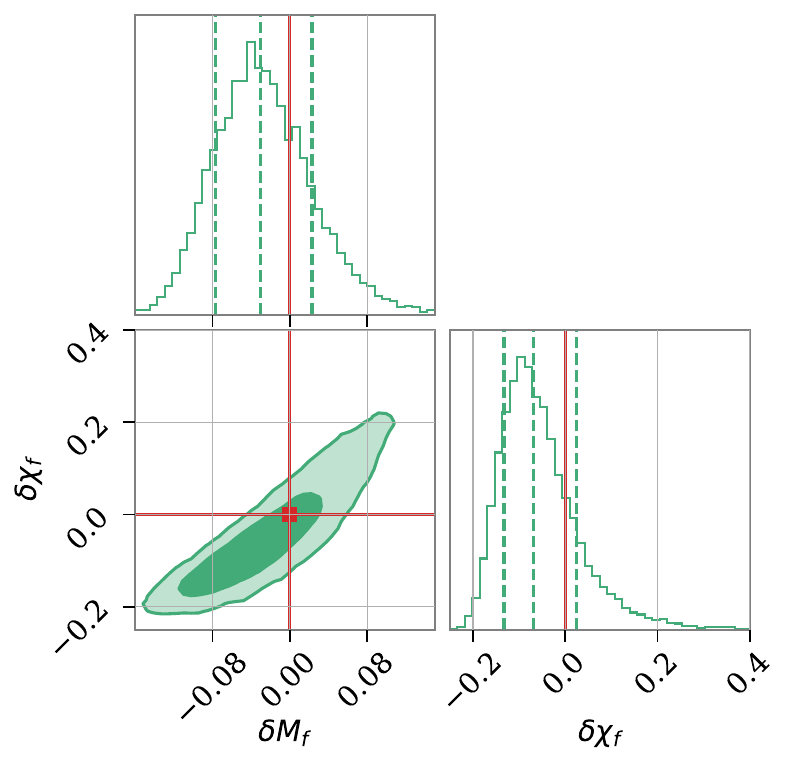}
  \caption{Posterior distributions of the fractional deviation parameters.
  The dark and light green contours enclose the $68\%$ and $95\%$ credible regions, respectively.
  The red lines indicate the null values corresponding to exact agreement with the conjecture. 
  The dashed lines denote medians and $68\%$ credible intervals. 
  }
  \label{fig:delta_mf-chif}
\end{figure}
Fig.~\ref{fig:delta_mf-chif} shows their joint and marginalized posterior distributions.
The two deviations are also correlated, inheriting the mass-spin degeneracy of the ringdown analysis. The posterior medians and $68\%$ credible intervals are
\begin{equation}
\delta M_f = -0.03_{-0.05}^{+0.05},
\qquad
\delta\chi_f = -0.07_{-0.06}^{+0.09},
\end{equation}
Both marginal distributions peak at slightly negative values. The null point lies inside the most concentrated part of the posterior. We therefore find no clear inconsistency between the remnant predicted by the maximum-entropy conjecture and the remnant measured from the ringdown of GW250114.


Because the conjecture concerns the entropy of the final black hole, 
we additionally compare the Kerr entropy constructed from the two remnant posteriors. 
\begin{figure}[ht]
  \centering
  \includegraphics[width=\columnwidth]{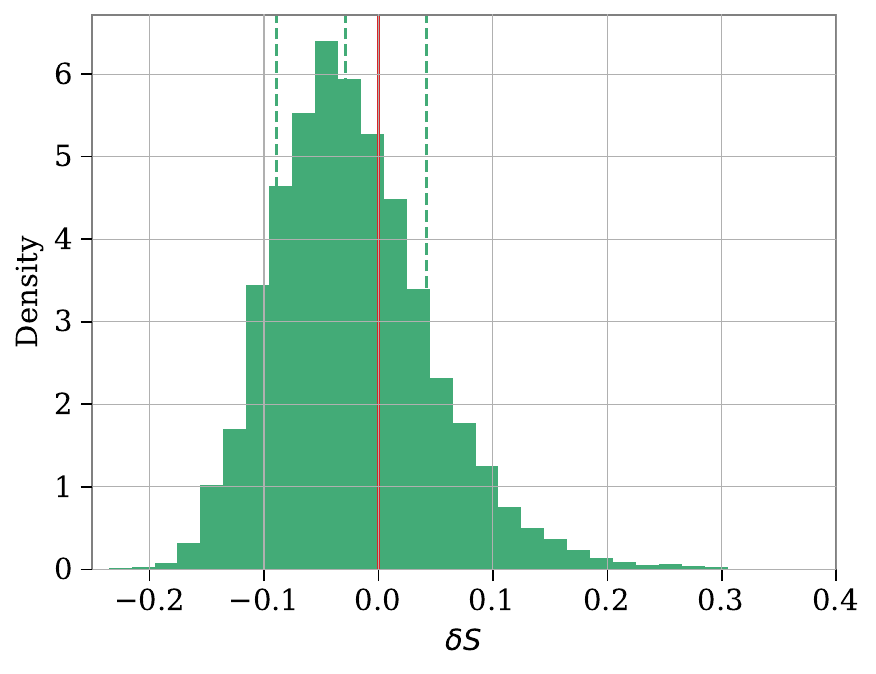}
  \caption{Posterior distribution of the fractional entropy deviation. 
  The red vertical line denotes the null value and the dashed lines denote the median and the $68\%$ credible interval.}
  \label{fig:delta_S}
\end{figure}
\Fig{fig:delta_S} shows the posterior distribution of $\delta S$ for the fiducial analysis. 
It projects the joint discrepancy of $M_f$ and $\chi_f$ onto the thermodynamic quantity that directly motivates the conjecture.
The distribution is mildly asymmetric, with a longer tail toward positive values.
Its median is $-0.03$, while the central $68\%$ interval extends from approximately $-0.09$ to $0.04$. 
The null value $\delta S=0$ lies well within the $68\%$ credible interval.
Thus, although the posterior maximum and median are slightly displaced from zero, the displacement is small compared with the statistical uncertainty and does not constitute evidence for a violation of the conjecture.

The separation of a GW signal into pre-merger and post-merger portions is not unique. 
Moreover, the validity of a linear QNM description depends on how long after the strain peak the ringdown analysis is started. 
To test the robustness of our result, we repeat the analysis for pre-merger truncation times.
\begin{figure}[ht]
  \centering
  \includegraphics[width=\columnwidth]{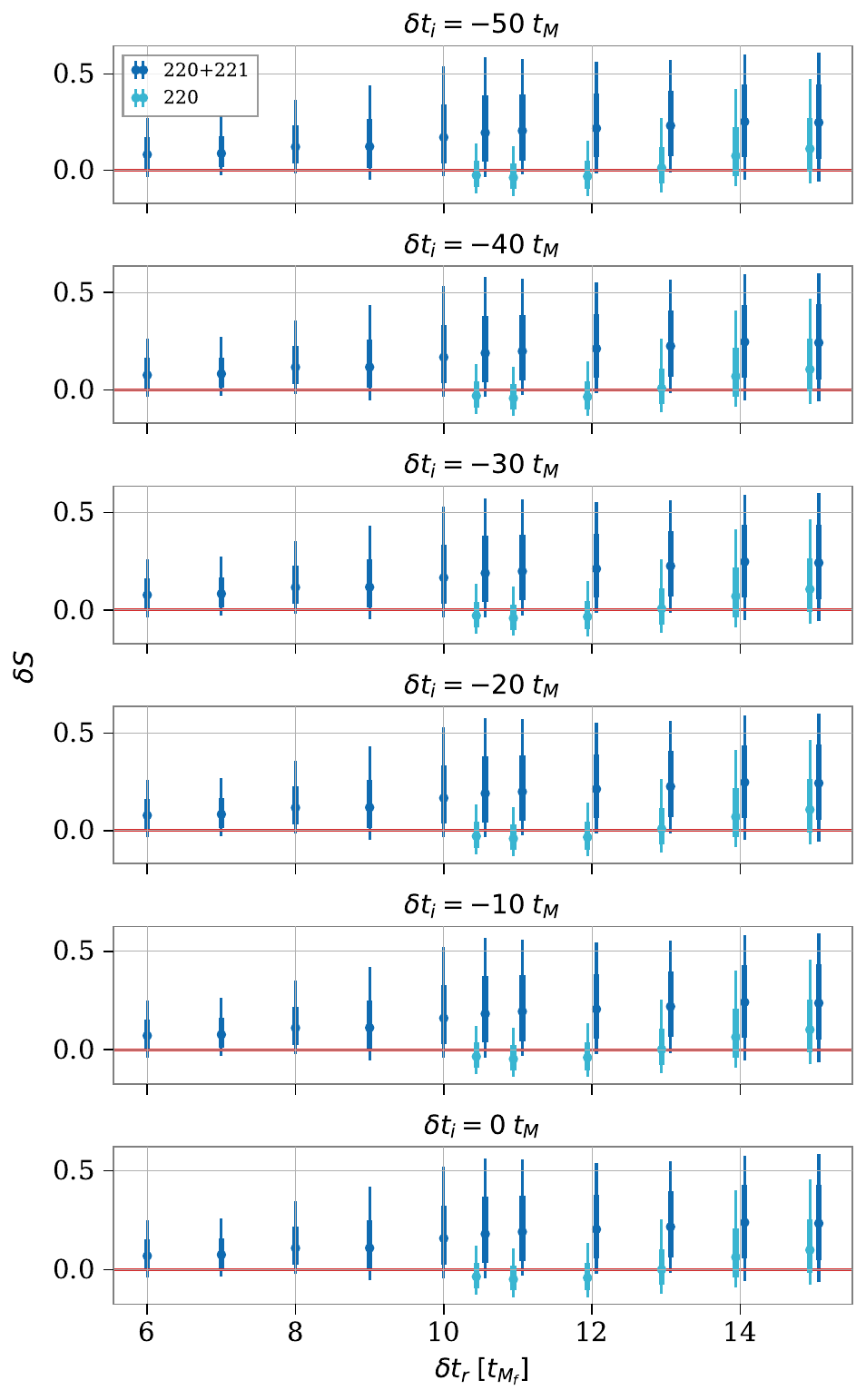}
  \caption{Dependence of the entropy-deviation posterior on the pre-merger truncation time, ringdown start time, and QNM content.
  The horizontal axis gives the ringdown start time in units of $t_{M_f}$.
  Dark-blue points show results obtained with the fundamental $220$ mode and its first overtone $221$, 
  while light-blue points show results using only the fundamental 220 mode.
  The points denote the posterior medians, and the thick and thin error bars represent the $68\%$ and $95\%$ credible intervals, respectively.
  The red horizontal line marks $\delta S=0$.}
  \label{fig:delta_S_array}
\end{figure}
\Fig{fig:delta_S_array} presents the medians and credible intervals of $\delta S$ for these analysis configurations. 
The thick and thin error bars denote the $68\%$ and $95\%$ credible intervals, respectively.
For a fixed ringdown model and start time $\delta t_{\rm r}$,
varying the pre-merger truncation time produces only a small change in the median of $\delta S$. 
The nearly identical behavior across the six panels indicates that the maximum-entropy prediction is stable against the amount of late-inspiral data included in the pre-merger analysis.

A larger dependence is associated with the post-merger analysis. 
For the $220+221$ model, the median entropy deviation is slightly positive at the earliest start times and generally increases as the start time is moved later. 
At the same time, the credible intervals broaden substantially because the ringdown amplitude decays rapidly and progressively less signal is retained. 
The increasing median at late start times should therefore not be interpreted as evidence for a growing physical deviation; 
it is accompanied by a loss of constraining power.
For start times at which both ringdown models are considered, 
the fundamental mode only analysis generally gives values of $\delta S$ closer to zero, 
and at several start times its median is slightly negative.
Adding the first overtone shifts the median toward positive values.
This difference illustrates the correlation between the assumed QNM content and the recovered remnant parameters.
Nevertheless, the $95\%$ credible intervals obtained with both mode choices remain compatible with $\delta S=0$ over the explored range of start times. 
The fundamental-mode results are also generally consistent with zero within their $68\%$ intervals, 
whereas the two-mode model shows a modest positive offset for some configurations.
Taken together, these tests show that the main conclusion is not driven by a particular pre-merger truncation time,
and none of the configurations considered yields a statistically significant entropy mismatch.  

\section{Conclusion}
In this work, we have tested the recently proposed maximum-entropy conjecture for binary black-hole mergers using GW250114. 
The conjecture states that the remnant BH corresponds to the Kerr configuration that maximizes the entropy along the mass--angular-momentum evolution of the binary. 
Using the 4PN energy--angular-momentum relation for a nonspinning, quasicircular binary, 
we mapped the pre-merger posterior of GW250114 into a prediction for the final mass, spin, and Kerr entropy. 
We then compared this prediction with an independent measurement of the remnant obtained from a ringdown analysis.

The maximum-entropy construction predicts a remnant concentrated around $M_f^{\rm i}\simeq70.0\,M_\odot$ and $\chi_f^{\rm i}\simeq0.70$. 
This prediction overlaps the high-probability region of the ringdown posterior, 
although the latter is substantially broader and exhibits a strong mass--spin correlation. 
The null values $\delta {M_f}=\delta {\chi_f}=\delta S=0$ remain compatible with the measurements at the $68\%$ credible level. 
We therefore find no statistical evidence for a violation of the maximum-entropy conjecture.

We also investigated the robustness of this conclusion by varying the termination time of the pre-merger analysis, 
the starting time of the ringdown analysis, and the QNM content of the ringdown model. 
The maximum-entropy prediction changes only weakly when the pre-merger truncation time is varied between $-50\,t_M$ and the reference merger time. 
The result is more sensitive to the post-merger start time and to whether the ringdown model contains only the fundamental $220$ mode or additionally includes the first overtone $221$. 
Later start times generally lead to broader posteriors because progressively less ringdown signal is retained, while including the overtone tends to shift the median entropy deviation toward positive values. 
Despite these variations, all of the analysis configurations considered remain consistent with $\delta S=0$.

Our results show that GW250114 is consistent with the maximum-entropy conjecture within the present statistical and waveform model uncertainties. 
This agreement should not be interpreted as a confirmation of the conjecture: the precision of the test is currently limited by the ringdown measurement, 
and the formulation adopted here relies on a nonspinning 4PN relation applied to a source whose component spins are small but not known to vanish exactly. 

Future observations with stronger ringdown signals will substantially improve this test. 
A more general treatment should incorporate spin-dependent PN dynamics, precession, higher-order corrections, and a systematic marginalization over the merger time and QNM content. 
Combining information from a population of sufficiently loud BBH events will further distinguish a genuine thermodynamic relation from event-specific statistical fluctuations. 
Such developments will make it possible to determine whether entropy maximization is merely a remarkable approximation to the result of NR or reflects a deeper principle governing the final state of black-hole mergers.

\section*{Acknowledgments}
YJ is supported by the China Postdoctoral Science Foundation under Grant Number 2025M783376. 
QGH is supported by the grants from NSFC (Grant No.~12547110, 12475065, 12447101).

\bibliography{refs.bib}
\end{document}